\documentclass[twoside,twocolumn,9pt]{article}
\usepackage{extsizes}
\usepackage[super,sort&compress,comma]{natbib}
\usepackage[version=3]{mhchem}
\usepackage[left=1.5cm, right=1.5cm, top=1.785cm, bottom=2.0cm]{geometry}
\usepackage{balance}
\usepackage{mathptmx}
\usepackage{sectsty}
\usepackage{graphicx}
\usepackage{lastpage}
\usepackage[format=plain,justification=justified,singlelinecheck=false,font={stretch=1.125,small,sf},labelfont=bf,labelsep=space]{caption}
\usepackage{float}
\usepackage{fancyhdr}
\usepackage{fnpos}
\usepackage[english]{babel}
\addto{\captionsenglish}{%
  
}
\usepackage{array}
\usepackage{droidsans}
\usepackage{charter}
\usepackage[T1]{fontenc}
\usepackage[usenames,dvipsnames]{xcolor}
\usepackage{setspace}
\usepackage[compact]{titlesec}
\usepackage{hyperref}

\usepackage{epstopdf}

\definecolor{cream}{RGB}{222,217,201}

\begin{document}

\pagestyle{fancy}
\thispagestyle{plain}
\fancypagestyle{plain}{
\renewcommand{\headrulewidth}{0pt}
}

\makeFNbottom
\makeatletter
\renewcommand\LARGE{\@setfontsize\LARGE{15pt}{17}}
\renewcommand\Large{\@setfontsize\Large{12pt}{14}}
\renewcommand\large{\@setfontsize\large{10pt}{12}}
\renewcommand\footnotesize{\@setfontsize\footnotesize{7pt}{10}}
\makeatother

\renewcommand{\thefootnote}{\fnsymbol{footnote}}
\renewcommand\footnoterule{\vspace*{1pt}%
\color{cream}\hrule width 3.5in height 0.4pt \color{black}\vspace*{5pt}}
\setcounter{secnumdepth}{5}

\makeatletter
\renewcommand\@biblabel[1]{#1}
\renewcommand\@makefntext[1]%
{\noindent\makebox[0pt][r]{\@thefnmark\,}#1}
\makeatother
\renewcommand{\figurename}{\small{Fig.}~}
\sectionfont{\sffamily\Large}
\subsectionfont{\normalsize}
\subsubsectionfont{\bf}
\setstretch{1.125} 
\setlength{\skip\footins}{0.8cm}
\setlength{\footnotesep}{0.25cm}
\setlength{\jot}{10pt}
\titlespacing*{\section}{0pt}{4pt}{4pt}
\titlespacing*{\subsection}{0pt}{15pt}{1pt}

\fancyfoot{}
\fancyfoot[LO,RE]{\vspace{-7.1pt}\includegraphics[height=9pt]{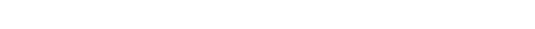}}
\fancyfoot[CO]{\vspace{-7.1pt}\hspace{13.2cm}\includegraphics{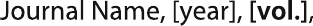}}
\fancyfoot[CE]{\vspace{-7.2pt}\hspace{-14.2cm}\includegraphics{head_foot/RF}}
\fancyfoot[RO]{\footnotesize{\sffamily{1--\pageref{LastPage} ~\textbar  \hspace{2pt}\thepage}}}
\fancyfoot[LE]{\footnotesize{\sffamily{\thepage~\textbar\hspace{3.45cm} 1--\pageref{LastPage}}}}
\fancyhead{}
\renewcommand{\headrulewidth}{0pt}
\renewcommand{\footrulewidth}{0pt}
\setlength{\arrayrulewidth}{1pt}
\setlength{\columnsep}{6.5mm}
\setlength\bibsep{1pt}

\makeatletter
\newlength{\figrulesep}
\setlength{\figrulesep}{0.5\textfloatsep}

\newcommand{\topfigrule}{\vspace*{-1pt}%
\noindent{\color{cream}\rule[-\figrulesep]{\columnwidth}{1.5pt}} }

\newcommand{\botfigrule}{\vspace*{-2pt}%
\noindent{\color{cream}\rule[\figrulesep]{\columnwidth}{1.5pt}} }

\newcommand{\dblfigrule}{\vspace*{-1pt}%
\noindent{\color{cream}\rule[-\figrulesep]{\textwidth}{1.5pt}} }

\makeatother

\twocolumn[
  \begin{@twocolumnfalse}
{\includegraphics[height=30pt]{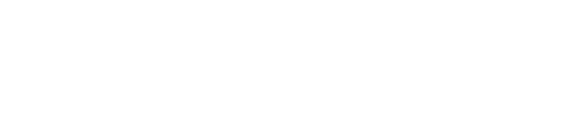}\hfill\raisebox{0pt}[0pt][0pt]{\includegraphics[height=55pt]{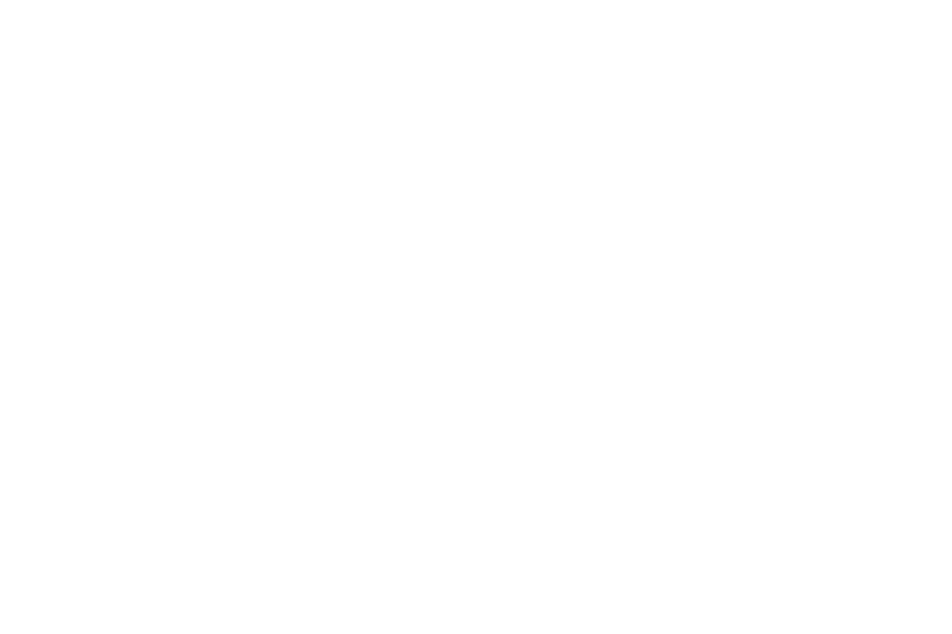}}\\[1ex]
\includegraphics[width=18.5cm]{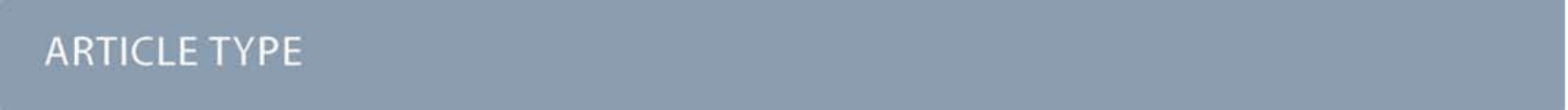}}\par
\vspace{1em}
\sffamily
\begin{tabular}{m{4.5cm} p{13.5cm} }

\includegraphics{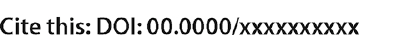} & \noindent\LARGE{\textbf{Mesoscale Simulation Approach for Assembly of Small Deformable Objects}} \\
\vspace{0.3cm} & \vspace{0.3cm} \\

 & \noindent\large{Toluwanimi O. Bello, Sangwoo Lee, and Patrick T. Underhill\textit{$^{\ddag}$}} \\

\includegraphics{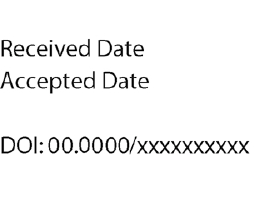} & \noindent\normalsize{We adapt Vertex models to understand the physical origin of the formation of long-range ordered structures in repulsive soft particles. The model incorporates contributions from the volume and surface area of each particle. Sampling using Monte Carlo simulations allows the system to naturally select preferred structures. We observe transitions between a body-centered cubic ordered state and a disordered state. Constraints to the simulation domain can suppress or allow the system to follow a path similar to Martensitic transformations from one ordered state to another ordered state. Finally, we show that rapid quenches from a disordered state into the ordered region lead to metastable local particle arrangements instead of a large-scale single crystal.}

\end{tabular}

 \end{@twocolumnfalse} \vspace{0.6cm}

  ]

\renewcommand*\rmdefault{bch}\normalfont\upshape
\rmfamily
\section*{}
\vspace{-1cm}


\footnotetext{Chemical and Biological Engineering, Rensselaer Polytechnic Institute, Troy, NY, USA.}

\footnotetext{\ddag~E-mail: underp3@rpi.edu}


\section{Introduction}
Material properties are directly influenced by their packing structures. Many studies of packing structures use inter-particle potentials to study the preferences among structures\cite{marcotte2011optimized}. However, if the potentials are not derived from a physical process, it can be difficult to gain physical insight into the origins of packing structures. For hard particles (either spherical or non-spherical), the packing structures are related to the volume fraction of particles. However, many objects are deformable including gas bubbles in foams \cite{van2006crystal,gabbrielli2012experimental,meagher2011analysis}, liquids drops in emulsions\cite{lacasse1996model,lissant1966geometry}, microgel particles \cite{seth2011micromechanical,meeker2004slip}, surfactant or block co-polymer micelles\cite{lee2014sphericity,fischer2011colloidal}, biological cells\cite{bi2016motility,derenyi2002formation}, and soft colloids \cite{batista_miller_2010,stadik2021deformable}. For such objects, the packing structures also depend on the deformability of the objects. This relationship is the subject of this article. We focus on systems that are so concentrated that there are nearly no gaps between the objects. Volume fraction is no longer an important quantity since it is nearly one. This also represents systems like block co-polymer melts for which the single component fills all space.

A common model for structures that fill all space is a vertex model. For example, these have been used to study growth of crystal grains {\cite{syha_weygand_2009}}, foams in the dry limit {\cite{ritacco_2020}}, and packings of cells {\cite{alt_ganguly_salbreux_2017,merkel_manning_2018}}. A vertex model divides space into polyhedra in 3D which are determined by the locations and connectivity of the vertices. The ``energy'' of the system is determined by the sizes and shapes of the polyhedra. Since the polyhedra fill all space, the shapes are coupled, leading to effective interactions between objects. Because the polyhedra are often composed of self-assembled constituents, the ``energy'' represents a free energy of the constituents within the constraints imposed by the polyhedra.

The relationship between energy and shape depends on the type of system. The Kelvin problem considers minimizing interfacial contact, so would represent the ground state of a system with energy based on interfacial contact\cite{weaire1997kelvin}. The Quantizer problem considers minimizing the second moment of the mass distribution of each object, so would similarly represent the ground state of a system with energy based on this second moment\cite{klatt_2019}.

For diblock co-polymers, self-consistent field theory (SCFT) or the diblock foam model (DFM) can be used to estimate this energy for a set of prescribed perfect crystal structures\cite{lee2013fluctuations,olmsted1998strong,olvera1991transitions,grason2005self}. These approaches can compare different crystal structures, but the structures are typically prescribed and translational entropy of the objects is not included. This also means that order-disorder transitions cannot be described since the disordered state is not in a prescribed crystal structure and has a significant contribution from translational entropy. One approach to overcome this is to approximate the entropic part analytically. Current approximations include the available region accessible per object, but do not account for correlations between the objects\cite{ziherl2000soap,ziherl2001maximizing}.

An approach based on the Quantizer problem has been simulated using Molecular Dynamics (MD) and Monte Carlo (MC) algorithms which directly sample the translational degrees of freedom and entropy\cite{hain2020low}. An order-disorder transition is found between a body centered cubic (BCC) arrangement and a disordered liquid-like state. Order-disorder transitions have also been seen in other soft models and potential-based models. However, the crucial difference between our work and many soft particle models is that our model produces interactions that are not pairwise-additive.

In this article, we use an approach based on the Kelvin problem. We also incorporate the principle that objects can exchange constituents which makes the objects have different sizes. This is observed in Frank-Kasper phases, so must be included in a model used to understand the formation of those phases\cite{reddy2018stable}. We use
Monte Carlo (MC) simulations to sample the equilibrium behavior of the geometry-based soft particle model in three dimensions. The particles fill all space, which requires them to deform based on the other particles. Importantly, we do not impose a crystal structure like is typically done in SCFT, but determine the equilibrium distribution of configurations through the simulation.

We first describe the vertex model which quantifies the energy of a configuration based on the shapes, along with the method used for simulation and quantities used to quantify the order. We then describe the behavior when each object has nearly the same size. This model produces an order-disorder transition between a BCC arrangement and a disordered liquid-like state. We then show the importance of the boundary condition imposed by the shape of the periodic simulation. Finally, we quantify the response to quenches from the disordered phase to the ordered region of the phase diagram.

\section{Model and Simulation Methods}

\subsection{Vertex Model}
The deformable self-assembled objects in the model fill all space with no gaps. We assume that they form polyhedra in three dimensions which are determined by the positions and connectivity of the vertices. The energy of the system is determined by the size and shape of the polyhedra. Because the objects fill all space, their shapes are coupled leading to effective interactions that are not pairwise additive. Specifically, the energy of $N_p$ particles that completely fill the volume $V$ is
\begin{equation}\label{eq:E3Dformula}
E =\sum_{i=1}^{N_p} \left( \frac{1}{2} k_3 (V_i-V_0 )^2+\gamma_3 (S_i-S_{si} ) \right) ,
\end{equation}
where $V_i$ and $S_i$ are the volume and surface area respectively of the $i$th polyhedron, $V_0$ is the preferred volume of a self-assembled particle, $S_{si}=6^{2/3} \pi^{1/3} V_i^{2/3}$ is the surface area of a sphere with the same volume as the $i$th polyhedron, $k_3$ is the parameter that determines the energetic cost to volume changes in 3D, and $\gamma_3$ is the parameter (like a surface tension) that determines the energetic cost to surface area changes in 3D.

The first term in the energy equation penalizes a particle from having a volume different from its preferred volume. Since $V_0$ is defined as the volume with minimum energy, we expect the energy to be quadratic for small deviations from that volume similar to the elastic energy of a Hookean spring. For large $k_3$, states with particles that vary in volume will have a large energy. This is appropriate for cells that cannot exchange material. At the other extreme are emulsions and foams where the dispersed phase does not directly prefer objects with a certain volume; $k_3$ will be near zero. Self-assembled micelles from surfactants or block co-polymers typically have a preferred aggregation number that sets a preferred volume. There is a free energy cost to deviating from this size. In the coarse-grained model, this is represented as the energy cost of the configuration. Instead of mapping a particular polymer or surfactant chemistry to a value of $k_3$, in this article we focus on the general features of the phase diagram for relatively large $k_3$. A similar type of energetic term occurs in other vertex models where its used to impose equal size objects as a soft constraint \cite{farhadifar2007influence,manning2010coaction,hufnagel2007mechanism}.

The second term in the energy equation quantifies the difference in energy between the object with surface area $S_i$ and a hypothetical spherical object with volume $V_i$. Inspired by the Kelvin problem, we take the energy difference to be a parameter $\gamma_3$ times the change in surface area. This area difference can be written using the isoperimetric quotient IQ as
\begin{equation}
S_i-S_{si}=6^{2/3} \pi^{1/3} V_i^{2/3} (IQ^{-1/3}-1)
\end{equation}
where IQ=$36 \pi V^{2}/S^{3}$ with V and S being the volume and surface area of the particle respectively. For equal volume particles, the energy in Equation~\ref{eq:E3Dformula} becomes
\begin{equation}
E \to \textrm{constant} + \gamma_3 \sum_{i=1}^{N_p} S_i,
\end{equation}
so the smallest interfacial surface area gives the lowest energy state (as in the Kelvin problem). In similar models with cellular tissues a preferred shape that is not spherical is used because of binding of cells \cite{bi2015density,bi2016motility,merkel_manning_2018}. Here, we assume that the preferred shape is spherical. The quantizer problem mentioned in the previous section also computes an energy with a minimum for a spherical shape, but with a slightly different way to quantify the deviation from spherical\cite{hain2020low}.

For packings of micelles, the micelles can exchange constituents to produce non-equal volumes that are necessary in the crystal structures with non-equivalent lattice site volumes such as Frank Kasper phases. In this general case, the energy of each object is written relative to a hypothetical reference state where each object is a sphere with volume $V_0$. The first term in equation~\ref{eq:E3Dformula} is the change in energy from the reference state to a sphere of volume $V_i$. The second term in equation~\ref{eq:E3Dformula} is the change in energy when the volume is fixed at $V_i$ and the shape changes from a sphere with surface area $S_{si}$ to the polyhedron with surface area $S_i$.

The model will be simulated using dimensionless variables, denoted with tildes. We scale energies by the thermal energy $k_B T$ where $k_B$ is Boltzmann's constant and $T$ is the absolute temperature. Therefore, we define $\tilde{E}=E/(k_B T)$. We scale lengths by $V_{0}^{1/3}$. This leads to two key dimensionless variables: $\tilde{k}_3 = k_3 V_0^2 /(k_B T)$ and $\tilde{\gamma}_3 = \gamma_3 V_0^{2/3} /(k_B T)$. Using these variables, the energy of the system is
\begin{equation}\label{eq:Etilde3Dformula}
\tilde{E} =\sum_{i=1}^{N_p} \left( \frac{1}{2} \tilde{k}_3 (\tilde{V}_i-1 )^2+\tilde{\gamma}_3 (\tilde{S}_i-\tilde{S}_{si} ) \right) .
\end{equation}

The final parameter in the system is the dimensionless total volume $\tilde{V} = V/V_0$. Because the particles fill all space, this is also equal to the sum of the particle volumes as $\tilde{V}=\sum_{i=1}^{N_p} \tilde{V}_i$. All simulations in this article consider the case where $\tilde{V}=N_p$. Because the total dimensionless volume equals the number of particles, the average dimensionless particle volume is one. Since the energy function contains $(\tilde{V}_i-1 )^2$, this means that the average dimensionless particle volume is the same as the preferred volume (the volume with the lowest energy).

\subsection{Monte Carlo Simulations}
As shown previously, the lowest energy state of the system with equal volume objects is the state with the smallest interfacial surface area. At non-zero temperature, the system will sample states according to their effective free energy including the effects of translational entropy. In this article, we sample this distribution using the Monte Carlo simulation technique at constant number of particles and constant temperature \cite{frenkel2001understanding}. Because the particles are self-assembled, the number of particles does not have to be constant; particles could break apart or fuse together. In this article, we fix the number of particles, so consider the case where the energy barriers to changing the number of particles is high enough such that it effectively does not happen.

Other than the number of particles, the two parameters that determine the system are $\tilde{k}_3$ and $\tilde{\gamma}_3$. Because temperature is used in the nondimensionalization, these parameters quantify how much thermal fluctuations can exchange volume between particles and can create more interfacial area. Here we focus on $\tilde{k}_3=1000$. From the simulations, thermal fluctuations lead to approximately $30\%$ variation in volume among the particles. This keeps each particle with nearly the same volume and allows us to compare with other models that focused on equal volume objects. Larger values of $\tilde{k}_3$ further restrict variations of volume, which increases the energy barriers for particle movement and reduces the computational efficiency of sampling phase space.

For this case where each particle has nearly the same volume, we expect the contact surface between two particles will be nearly equidistant from the particle centers. That is, we expect the particle shapes to form the regions of a Voronoi tessellation {\cite{10.1145/235815.235821}}. This approximation has been used previously in models of cellular tissues. For each particle configuration, the state is determined by the positions of the Voronoi centers. To compute the energy, the Voronoi tessellation is computed to determined the positions and connectivity of the vertices. These are used to calculate the volume and surface area of each particle, which determines the energy of the configuration.

The standard Metropolis algorithm is used to sample the equilibrium states. After a change to the configuration is performed, the change in energy is computed. If the energy of the system goes down, the change is accepted. If the energy of the system goes up, the move is accepted with probability $\exp (-\Delta \tilde{E})$. In simulations with a fixed simulation box, each step consists of randomly choosing one particle and randomly displacing the center of its Voronoi cell. The maximum displacement is chosen to keep the acceptance rate in the range $30\% - 60\%$. A Monte Carlo cycle consists of $N_p$ steps so that each particle is chosen once on average per Monte Carlo cycle.

Some simulations are performed in which the shape of the simulation box can fluctuate. This would be analogous to simulations at constant pressure instead of constant volume. In simulations at constant pressure, a term $P \Delta V$ would be included in the exponential of the Metropolis algorithm. However, we consider the whole system to be nearly incompressible. Therefore, the shape of the box must change to maintain constant volume. This is done by randomly choosing a coordinate direction and randomly choosing the amount to scale (stretch or compress) that direction. The other two directions are scaled equally to the amount needed to keep the volume constant. All particle centers are moved together affinely within the simulation box. The maximum amount of scaling allowed is chosen such that the acceptance rate is in the range $30\% - 60\%$.

In simulations with a fluctuating simulation box, a simulation step starts by randomly deciding whether that step consists of scaling the box or moving a single particle. The probability of scaling the box is $1/N_p$ so that one box scaling occurs on average per Monte Carlo cycle. This is done because a box scaling moves all particles affinely, while a particle move only displaces a single particle.

For a typical ``melt'' simulation, the simulation starts at a specific $\tilde{k}_3$ and high $\tilde{\gamma}_3$ in a perfect crystal and is run for at least 1000 Monte Carlo cycles until it equilibrates at the initial $\tilde{\gamma}_3$. The simulation box is chosen to be cubic and commensurate with the starting crystal. The cubic shape is useful because it maximizes the smallest side length, which often determines the influence of periodic boundary conditions. Then, we take the end of that run and and use it as the initial condition for a smaller $\tilde{\gamma}_3$ and run that until equilibration. That continues until the smallest $\tilde{\gamma}_3$. Each condition is run for least 1000 MC cycles to allow for equilibration. Conversely, ``quench'' simulations start at a specific $\tilde{k}_3$ and low $\tilde{\gamma}_3$ with each subsequent increased $\tilde{\gamma}_3$ initialized by the end of the previous case, with a similar allowance for equilibration. For the quench simulations, we chose the simulation box to be commensurate with one of the investigated crystals to allow a single crystal to form if that was preferred by the system.

\subsection{Order Characterization}
We use multiple quantities to measure the order of the system. Here we define and summarize these measures. In the BCC packing structure, each deformable particle is a polyhedron with 14 sides/surfaces. This corresponds to the 14 neighbor particles that touch at those surfaces. In a disordered packing, the particles in the system will have a distribution of number of neighbors (sides of polyhedron). For each configuration of the system, we compute the sample standard deviation of the number of neighbors among the particles. Averaging this quantity over sampled configurations is denoted as $std_{nn}$. This is zero in an ordered packing structure where each particle has the same number of sides, and is positive when there are defects or the system is disordered.

Within ordered regions, we combine modifications of the Steinhardt order parameters proposed by Dellago et al.\cite{lechner2008accurate} and Mecke et al.\cite{mickel2013shortcomings}. For each particle, we compute
\begin{equation}\label{eq:orderparameterformula}
q_{lm}' =\sum_f \frac{A(f)}{A} Y_{lm} (\theta_f,\phi_f)  ,
\end{equation}
where the $f$ denotes the faces of the polyhedron corresponding to the neighbors, $Y_{lm}$ are the spherical harmonics, and $\theta$ and $\phi$ are spherical coordinate angle of the vector pointing from the particle towards its neighbor. The prime denotes that the spherical harmonics are weighted by the ratio of the area of a face to the total area of the particle.

Averaging over neighbor particles is useful for crystal identification in situations with thermal fluctuations using
\begin{equation}\label{eq:orderparameterformula}
\overline{q}_{lm}'(i) ={ \frac{1}{\tilde{N}_{n}(i)+1} \sum_{k=0}^{\tilde{N}_{n}(i)}{q}_{lm}'(k)} ,
\end{equation}
where $k$ is an index corresponding to the particle and all $N_n$ neighbor particles to particle $i$. Finally, this can be used to compute the order parameters using
\begin{equation}\label{eq:orderparameterformula}
\overline{q}_{l}'(i) =\sqrt{\frac{4\pi}{2l+1}\sum_{m=-l}^{l}|\overline{q}_{lm}'(i)|^2} .
\end{equation}
In our work we use both $\bar{{q}_4}'$ and $\bar{{q}_6}'$.

\section{Results and Discussion}
\subsection{Order/disorder transition with fixed simulation box}
\begin{figure}[h]
\centering
  \includegraphics[width=3.5in]{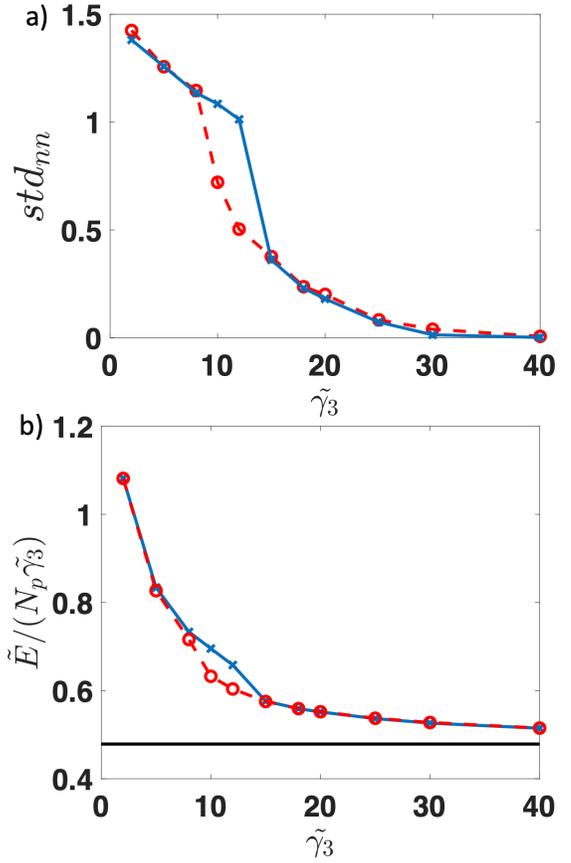}
  \caption{ Simulations of 128 particles at $\tilde{k}=1000$. (a) Order is quantified while increasing $\tilde{\gamma}_3$ (solid blue) and decreasing $\tilde{\gamma}_3$ (dashed red) through the order-disorder transition. (b) Quantification of the energy per particle per $\tilde{\gamma}_3$ with increasing $\tilde{\gamma}_3$ (solid blue) and decreasing $\tilde{\gamma}_3$ (dashed red) through the order-disorder transition. The horizontal solid black line denotes the value of the perfect BCC crystal.}
  \label{fgr:1}
\end{figure}

The first results that illustrate an order/disorder transition within the model considers the case of a fixed cubic simulation domain containing $128$ particles at $\tilde{k}=1000$. The quantities described previously are used to characterize the equilibrium behavior during a melt from the order to disorder state or a quench from the disorder to order state. Figure  \ref{fgr:1} shows the standard deviation of the number of neighbors and the scaled energy as a function of $\tilde{\gamma}_3$ across the transition. The perfect BCC structure without lattice fluctuations is $\tilde{E}=0.4788 N_p \tilde{\gamma}_3$. Therefore, we scale the energy so that it is a constant in the perfect BCC state. Note that because both $E$ and $\gamma_3$ were scaled by $k_B T$, the ratio is independent of $T$.

For $\tilde{\gamma}_3 \geq 15$, the melt and quench simulations produce the same ordered BCC structure with defects due to thermal fluctuations. These fluctuations give a state where not all particles have 14 neighbors and an energy larger than the energy of the perfect BCC structure. For $\tilde{\gamma}_3 \leq 8$, the melt and quench simulations produce the same disordered structures with even higher energy (but lower free energy because of the entropic contributions from translational degrees of freedom). For intermediate $\tilde{\gamma}_3$, the Monte Carlo simulations with different initial conditions remain in different metastable states for a large enough number of Monte Carlo cycles to produce different averages.

\begin{figure}[h]
\centering
  \includegraphics[width=3.5in]{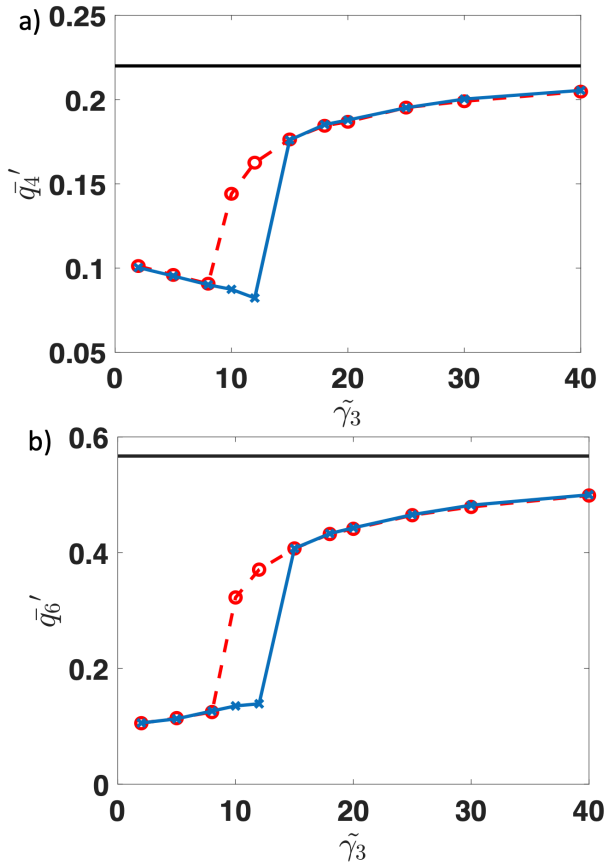}
  \caption{Simulations of 128 particles at $\tilde{k}=1000$. Order is quantified via order parameters $\bar{q}'_4$ (a) and $\bar{q}'_6$ (b) while increasing $\tilde{\gamma}_3$ (solid blue) and decreasing $\tilde{\gamma}_3$ (dashed red) through the order/disorder transition. The order parameters approach the perfect BCC values (solid black) for large $\tilde{\gamma}_3$.}
  \label{fgr:2}
\end{figure}

\begin{figure}[h]
\centering
  \includegraphics[width=3.5in]{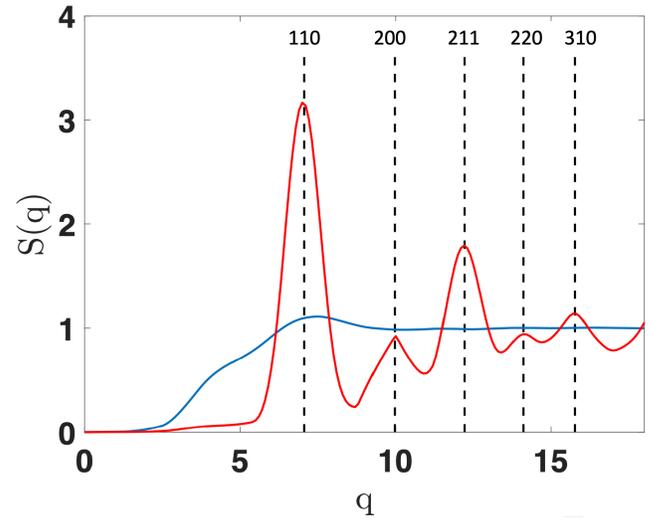}
  \caption{Simulations of 128 particles at $\tilde{k}=1000$. Order is quantified using structure factor at $\tilde{\gamma}_3=2$ (blue) and $\tilde{\gamma}_3=40$ (red). The structure factor at $\tilde{\gamma}_3=40$ matches the peak positions for a perfect BCC structure (dashed black).}
  \label{fgr:3}
\end{figure}

The order/disorder transition is also quantified using the order parameters in Figure \ref{fgr:2}. Both parameters have similar features in that they approach the BCC value for large $\tilde{\gamma}_3$, give smaller values for small $\tilde{\gamma}_3$, and show a transition within $8 < \tilde{\gamma}_3 < 15$. The presence of disorder and the BCC order are also confirmed using the three-dimensionally averaged structure factor on both sides of the transition in Figure \ref{fgr:3}. At $\tilde{\gamma}_3=40$, the peaks in the structure factor match with the theoretical positions for BCC. At $\tilde{\gamma}_3=2$, the system shows no long ranged order, with a liquid-like structure factor \cite{patel2017structure}. This liquid structure is likely the reason for the values and trends of the order parameters in Figure \ref{fgr:2} at low $\tilde{\gamma}_3 < 10$. Unlike order parameters in some other systems, the $\bar{q}'_4$ for low $\tilde{\gamma}_3$ does not approach zero. The reason for this is likely excluded volume interactions that produce some short-range order as neighbor particles pack together in the first-neighbor shell.

These Monte Carlo simulations naturally sample the thermal fluctuations that incorporate the translational entropy contributions to the free energy, and the resulting order/disorder transition. For simulations with small $\tilde{\gamma}_3$, the system reaches equilibrium in a state with energy higher than the minimum possible, but with a large number of configurations and low free energy. Similarly, the quench simulations show that the Monte Carlo simulations naturally produce the BCC structure from initial conditions that do not impose that structure.

Although our energy function is different, these results are consistent with work based on the Quantizer problem \cite{hain2020low}. For example, both systems show hysteresis within the transition range, a character of first-order transitions. The results are also similar in the sense that systems tend to have a preference for BCC formation on quenching from disordered states, while choosing a disordered state on melting.

\subsection{Order/disorder transition with fluctuating simulation box}
\begin{figure}[h]
\centering
  \includegraphics[width=3.5in]{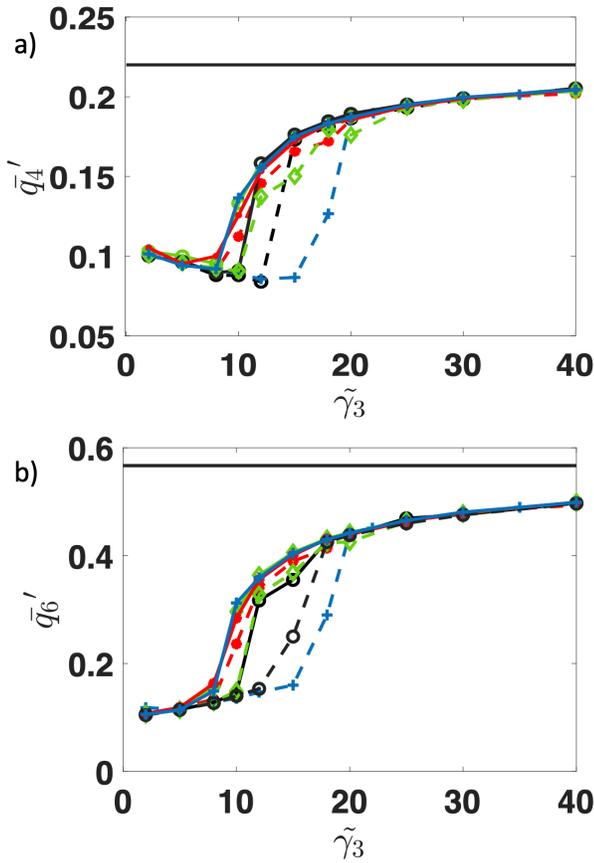}
  \caption{Order is quantified using $\bar{{q}_4}'$ (a) and $\bar{{q}_6}'$ (b) for simulations with $N_p$ of 54 (red asterisks), 72 (green diamonds), 96 (black circles), and 108 (blue crosses) particles at $\tilde{k}=1000$. For each $N_p$, the melt (decreasing $\tilde{\gamma}_3$) and quench (increasing $\tilde{\gamma}_3$) plot has the same color, but the melt is represented by solid lines, while the quench is represented by the dashed lines. The solid horizontal black lines represents the perfect BCC structure.}
  \label{fgr:4}
\end{figure}

We compared the melt and quench data for 54, 72, 96, and 108 particles in a fluctuating simulation box in order to measure the impact of particle number and box constraints on structure formation. Allowing fluctuations to the shape of the simulation box gives freedom to do the simulations with a finer resolution of number of particles while also allowing a single crystal to fit perfectly in the box. The order parameters from the simulations are shown in Figure \ref{fgr:4}.

The melt simulations start in BCC configuration in the ordered region and go through a order-disorder transition with decreasing $\tilde{\gamma}_3$. The quench simulations start in random disordered configurations and are quenched with increasing $\tilde{\gamma}_3$ into ordered states. The key variability among the simulations with different number of particles is in the transition region of the melt and quench.

For the systems with 54 or 72 particles, hysteresis in the transition is not observed. Within the constraints placed on the system by the periodic boundary conditions, it is easier to coordinate the particle positions to either melt or form a single crystal. Simulations with 96 particles show a small amount of hysteresis, while simulations with 108 particles show more hysteresis. The amount is comparable to or larger than earlier results with 128 particles. The additional allowable states in configuration space due to the box fluctuations may be the reason for the wider hysteresis region. These simulations use at least 1000 MC cycles at each parameter set during the quench. This is slow enough that the hysteresis loops show nearly no dependence on quench rate. We observe some dependence of the hysteresis width on the system size, but the details of the system make it difficult to determine the functional dependence. Only near 100 particles is the hysteresis measurable. Also, the need to have a particle number and box shape commensurate with a single crystal leads to only a few discrete cases where hysteresis can be quantified.

\begin{figure*}[h]
\centering
  \includegraphics[width=7in, trim = 0cm 7cm 0cm 6cm, clip]{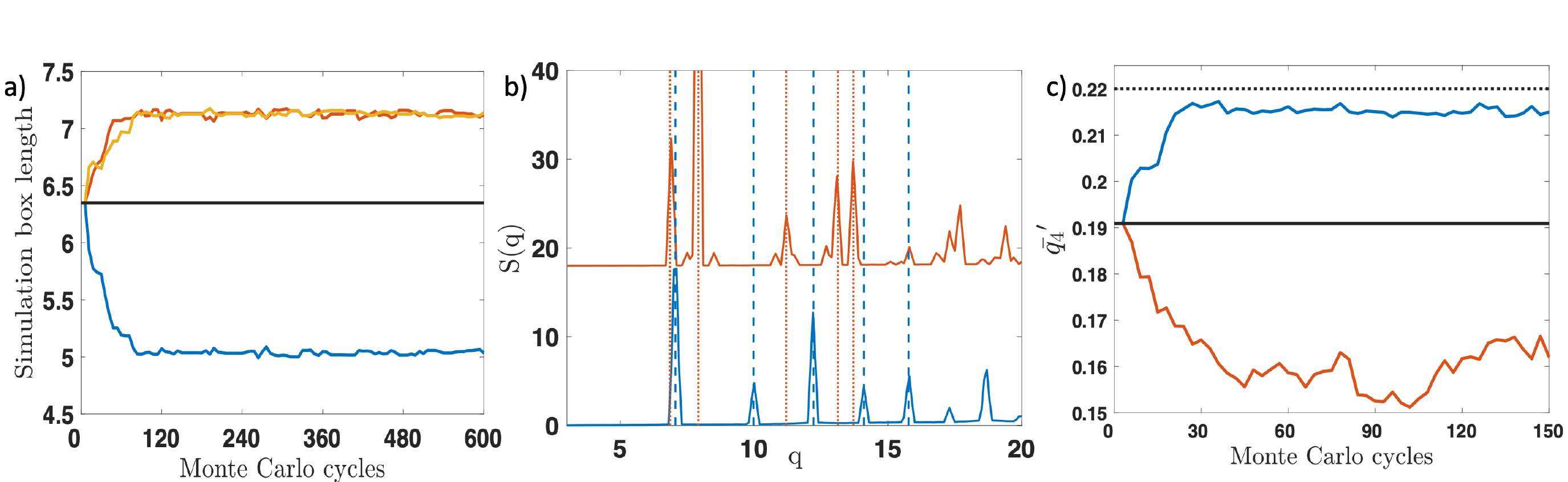}
  \caption{Simulations of 256 particles at $\tilde{k}=1000$ and $\tilde{\gamma}_3=40$ (a) Expansion of $x$ and $y$ dimensions (orange and yellow) and contraction of $z$ (blue) during the evolution from FCC to BCC configuration. The black line represents the initial cubic simulation box. (b) Order is quantified using structure factor to compare the final configuration which is BCC (blue) for the fluctuating case to that of the fixed case which stays FCC (red, shifted vertically for clarity). The dashed blue and dotted red lines are the peak positions for perfect BCC and FCC states respectively. (c) Order is quantified using $\bar{{q}_4}'$ to show the pathway from the initial FCC to BCC with thermal fluctuations (blue). The values for a perfect crystal are shown as dashed black for BCC and solid black for FCC. Simulations starting FCC in a fixed box case stays FCC with thermal fluctuations (red).}
  \label{fgr:5}
\end{figure*}

\subsection{Face-centered cubic (FCC) initial condition}
We have shown that with decreasing $\tilde{\gamma}_3$, BCC structures with various initial conditions eventually devolve into disordered states. BCC structures also spontaneously occur when quenching from disorder. However, for systems where the initial condition is FCC in a fluctuating simulation domain, we find a unique behavior in the ordered region. The simulations start FCC at $\tilde{\gamma}_3=40$ and begins to fluctuate over several Monte Carlo cycles once it is initialized. Eventually, at the same $\tilde{\gamma}_3=40$ the simulation domain expands in the $x$ and $y$-directions and contracts in the $z$-direction to ease the system into its preferred equilibrium state, as shown in Figure~\ref{fgr:5}a. The larger dimensions in this final preferred state are a factor of $\sqrt{2}$ larger than the smaller dimension. This ratio is important because it corresponds to a final state of BCC in a non-cubic simulation box. Note that the system randomly chooses which directions expand and contract. In other versions of the same simulation, other directions expand and contract, but consistently two sides expand and one contracts.

The final BCC structure is confirmed by looking at the structure factor in Figure \ref{fgr:5}b and the order parameter $\bar{{q}_4}'$ in Figure \ref{fgr:5}c. When going from FCC to BCC, the progression of the simulation follows a path similar to Martensitic transformations in that there is a cooperative movement of the particles within short distances to form the new state\cite{nishiyama2012martensitic}.

When we initialize the same system in a fixed simulation domain, it behaves differently. The system starting with a FCC structure remains with a FCC structure but with thermal fluctuations. This is confirmed by looking at the structure factor in Figure \ref{fgr:5}b and the order parameter $\bar{{q}_4}'$ in Figure \ref{fgr:5}c. This means that FCC is metastable and thermal fluctuations are not enough to change the structure in a fixed simulation box. This highlights that the constraints can influence what is observed.

\subsection{Metastability and equilibration}
\begin{figure}[hbt]
\centering
  \includegraphics[width=3.5in]{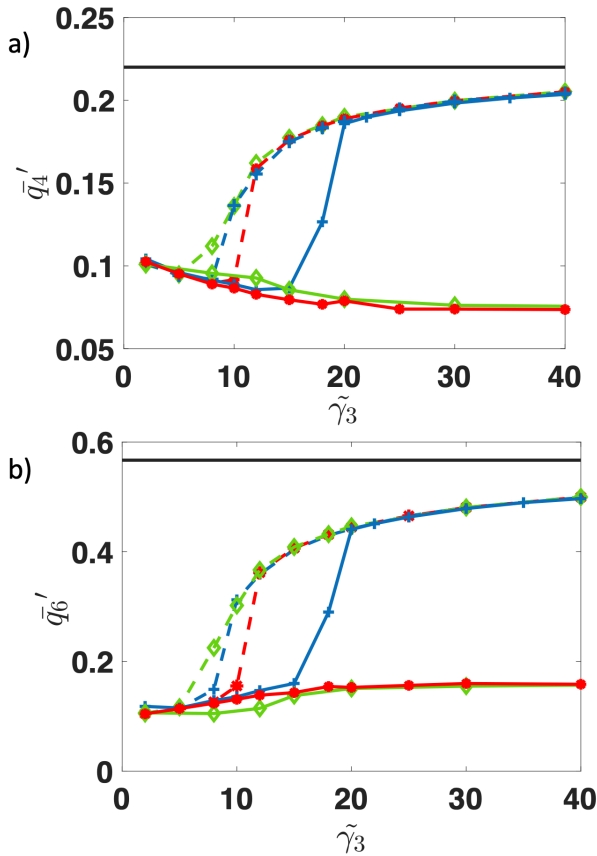}
  \caption{Order parameters $\bar{{q}_4}'$ (a) and $\bar{{q}_6}'$ (b) for simulations of 108 (blue crosses), 256(red asterisks), and 1024(green diamonds) particles at $\tilde{k}=1000$. The dashed lines are melts, while the same-colored solid lines are quenches. The solid horizontal black lines represents the perfect BCC structure.}
  \label{fgr:6}
\end{figure}

For quenches in systems with $N_p$ > 128 particles, there is a larger energy barrier to single crystal ordered structure formation. Figure \ref{fgr:6} shows the order parameters for simulations with 108, 256, and 1024 particles. Melt simulations starting in a BCC structure show a order-disorder transition seen previously with little dependence on the number of particles. The simulations with 108 particles were shown previously and have hysteresis in the transition but eventually reach a BCC structure during the quench simulations. The quench simulations of 256 and 1024 particles do not reach a single crystal BCC structure.

\begin{figure}[hb]
\centering
  \includegraphics[width=3.5in]{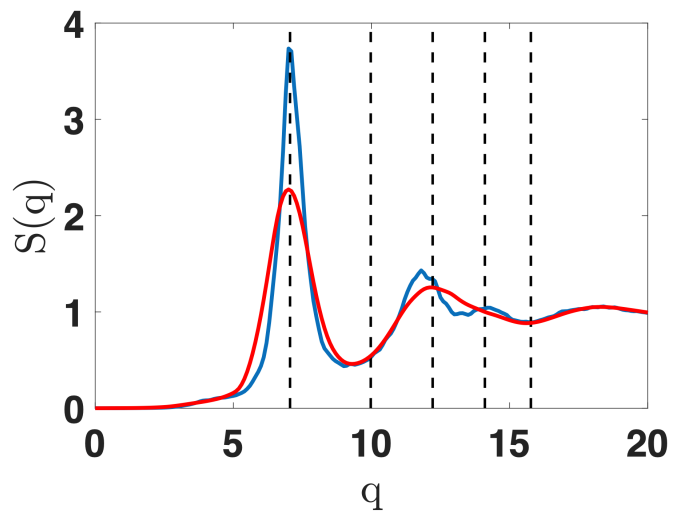}
  \caption{Simulations of 1024 (blue) and 108 (red) particles at $\tilde{k}=1000$. In both cases, there is an initial peak that corresponds to the shape of the system and there are barely any subsequent peaks. The broken black lines are the positions for the perfect BCC configuration.}
  \label{fgr:7}
\end{figure}

In order to test whether the particle arrangements for 256 and 1024 particles were do to the inability of the particles to arrange into the single crystal, we performed simulations with 108 particles but with a more rapid quench. If given enough time to equilibrate, these 108 particle simulations form a single BCC crystal (Figures \ref{fgr:4} and \ref{fgr:6}). Figure\ref{fgr:7} compares the structure factor for this rapid quench with the structure factor from the simulations with 1024 particles. The similarity, especially the lack of the BCC peak near a wavevector of 10, suggests that the local particle arrangements but lack of single crystal for 1024 particles is due to the barrier of the large systems to nucleate the BCC structure.

\section{Conclusions}
In this article, we used a modified version of previous vertex models to understand packings of soft particles. We showed a order-disorder transition driven by a balance between surface tension and thermal fluctuations. Our results are consistent with related work published by Ruscher, Baschnagel,
and Farago \cite{ruscher2016voronoi}, as well as Hain, Klatt and Schröder-Turk\cite{hain2020low} in that a slow quench from disordered configuration leads to the formation of BCC structure. Furthermore, when we compare melt and quench paths for the same system, we see hysteresis in the paths that are similar to those presented previously.

In addition to an order-disorder transition, we also studied the effects of simulation domain constraints on our results. Simulations starting in a FCC state in a fixed simulation box remain in a metastable FCC structure. However, when the simulation domain is allowed to fluctuate, the initial FCC configuration reorders itself to find a new preferred BCC state in a manner similar to a Martensitic transformation where particles move in a cooperative movement within short distances to form the new state. Finally, we showed that rapid quenches lead to local rearrangements of particles but that large-scale BCC structure is not able to nucleate and grow in those conditions.

This work has focused on a value of $\tilde{k}_3$ large enough that the particles have nearly equal volume. This simpler part of phase space is dominated by cubic structures. One important feature of Frank-Kasper phases is that there can be a variable in volumes among the particles due to exchange of material. The methodology used here can be used to understand the coupling of this exchange with the physical principle of sphericity and thermal fluctuations. The ability to simulate systems with many particles will also enable studies of structures that have been observed experimentally such as quasicrystals\cite{fischer2011colloidal, schwab1996thermotropic, vogt1994dynamics}. Future publications will look at parts of the phase space where Frank-Kasper phases and quasicrystals may occur. We also expect that future studies will use a multiscale approach, in which methods such as SCFT on a particular polymer will be used to determine the energy function for that molecule which will be used in the vertex model approach described here.

\section*{Conflicts of interest}
There are no conflicts to declare.



\balance



\providecommand*{\mcitethebibliography}{\thebibliography}
\csname @ifundefined\endcsname{endmcitethebibliography}
{\let\endmcitethebibliography\endthebibliography}{}

\end{document}